# Structure, magnetic properties, and cycloidal spin ordering in Nd-doped BiFeO$_3$ synthesized by sol-gel method.

K. Komędera[1,2], K. Siedliska[3], Ł. Gondek[1], D.M. Nalecz[2,4], R. Panek[5], M. Krupska-Klimczak[6,7], I. Jankowska-Sumara[7], W. Tokarz[1], W. Tabiś[1,8], T. Pikula[3*]

[1] *Faculty of Physics and Applied Computer Science, AGH University of Krakow, Mickiewicza 30, Kraków, 30-059, Poland*
[2] *Mössbauer Spectroscopy Laboratory, University of National Education Commission, Podchorążych 2, 30-084 Kraków, Poland*
[3] *Department of Electronics and Information Technologies, Lublin University of Technology, Nadbystrzycka 38A, 20-618 Lublin, Poland*
[4] *Institute of Art and Design, University of the National Education Commission, Podchorążych 2, 30-084, Kraków, Poland*
[5] *Department of Construction Materials Engineering and Geoengineering, Lublin University of Technology, Nadbystrzycka 40, 20-618 Lublin, Poland*
[6] *Institute of Security Studies and Computer Science, University of the National Education Commission, Podchorążych 2, 30-084 Kraków, Poland*
[7] *Department of Exact and Natural Sciences, University of the National Education Commission, Podchorążych 2, 30-084 Kraków, Poland*
[8] *Institute of Solid State Physics, TU Wien, 1040 Vienna, Austria*

*Corresponding author: *t.pikula@pollub.pl*

**Abstract**

In this study the structural and magnetic properties of Nd-doped BiFeO$_3$ materials with the formula Bi$_{1-x}$Nd$_x$FeO$_3$ ($x$ = 0–0.2) were investigated. Samples were prepared using the sol-gel method and annealed at 823 K for 4 hours. Microstructural studies indicated that nanomaterials with an average grain size of less than 100 nm were produced, with the smallest grain size observed for the solid solution with $x$ = 0.15. Rietveld analysis of diffraction patterns revealed a composition-driven transformation from the rhombohedral (*R3c*) lattice to the orthorhombic (*Pbam*) structure. The above finding is correlated to Mössbauer spectroscopy data, which demonstrated the gradual destruction of the spin cycloid due to the structural transformation to the *Pbam* structure. The spectra were analyzed using a complex hyperfine magnetic field distribution model reflecting cycloidal spin ordering. Scanning differential calorimetry revealed that for the investigated materials the magnetic ordering temperatures are close to 640 K, independently from the Nd concentration. Magnetic measurements showed the emergence of weak ferromagnetic properties with increasing Nd ions concentration, consistently with Mössbauer spectroscopy data.

# 1. Introduction

Multiferroic materials are a type of materials that display two or more spontaneous ferroic-type orderings at the same time within one material. These orderings can include (anti)ferroelectricity, (anti)ferromagnetism, and ferroelasticity. The coexistence of these properties within a single material volume makes multiferroic materials particularly intriguing [1]. They have a wide range of potential applications in fields such as sensors [2], data storage [3], spin-valve devices [4], actuators [5], spintronics [6], energy-harvesting devices [7], and many more. One of the multiferroic materials that has attracted significant attention in recent years is bismuth ferrite ($BiFeO_3$ or BFO), which belongs to the family of compounds called perovskites. It is a single-phase material with a rhombohedrally distorted perovskite structure and a polar space group *R*3*c* at room temperature. What sets BFO apart is its rare properties of the coexistence of antiferromagnetic nature with Néel temperature of $T_N$ ~ 643 K and ferroelectricity with Curie temperature of $T_C$ ~ 1103 K [8]. Hence, $BiFeO_3$ is treated as one of the top candidates for room-temperature magnetoelectric applications. However, cycloidally modulated magnetic structure with an incommensurate long-wavelength period of 620 Å hinders macroscopic magnetizations, which in turn reduces the linear magnetoelectric effect in bulk bismuth ferrite [9,10]. Another obstacle to its application is the large leakage current, which is mainly caused by the existence of secondary phases such as $Bi_2O_3$, $Fe_2O_3$, $Bi_2Fe_4O_9$, and $Bi_{25}FeO_{40}$. According to the complex phase diagram of $Bi_2O_3$-$Fe_2O_3$, the stability of single-phase BFO is limited to a very narrow range of temperatures, and the formation of stable secondary phases ($Bi_2Fe_4O_9$, and $Bi_{25}FeO_{40}$) is rather easy [11]. Therefore, obtaining phase pure materials of bismuth ferrites through conventional solid-state reactions requiring high sintering temperatures is quite challenging.

Various efforts have been made recently to overcome the disadvantages associated with pure phase preparation. These actions encompass: (1) synthesizing $BiFeO_3$ nanoparticles with a grain size smaller than 62 nm to attain ferromagnetic properties by disrupting the spiral spin structure; (2) proposing multiple alternative chemical synthesis methods to create a single-phase $BiFeO_3$, such as hydrothermal process, microwave synthesis, co-precipitation, combustion synthesis, sol-gel process, and more; (3) doping $ABO_3$ structure with foreign atom at either A or B site of the crystalline lattice. It has been noted that partial substitution of rare-earth elements like La, Nd, Pr, or Sm at the Bi site, even in small concentrations, prevents the formation of the secondary phases during the synthesis process. Moreover, Bi ions substitution leads to a strong structural change resulting from an imbalanced $6s^2$ lone pair of electrons and further affects intrinsic properties such as magnetic, dielectric, and optical responses [11].

This study aimed to replace Bi ions in the perovskite structure with varying concentrations of $Nd^{3+}$ ions, resulting in a solid solution of $BiFeO_3$ and $NdFeO_3$. This replacement is particularly interesting due to the magnetic properties of neodymium ferrite, which exhibits weak ferromagnetism attributed to a Dzyaloshinskii-Moriya interaction [12] with the Néel temperature $T_N$ = 690 K [13]. Increasing neodymium doping of BFO is accompanied by structural transitions, starting from the *R*3*c* phase for neodymium concentrations less than 0.1, and going through the mixing of two crystalline phases, *R*3*c* and *Pbnm*, for concentrations in the range from 0.15 to 0.45. Finally, under further doping the structure achieves the pure *Pbnm* phase for $x > 0.45$, which is typical for the $NdFeO_3$ compound [14]. Additionally, Nd doping reduced the number of oxygen vacancies resulting from bismuth volatilization during the synthesis process.

The $Bi_{1-x}Nd_xFeO_3$ (BNFO) solid solution has been the subject of numerous studies [14–20]. However, there is still a lack of information about the mechanism behind the destruction of the spin cycloid. The main objectives of this study were as follows: (1) to examine the gradual structural transformation resulting from increasing concentrations of Nd ions in the BiFeO$_3$ lattice, and (2) to investigate the properties of the spin cycloid and observe its destruction process. To achieve this, we synthesized a series of samples with varying levels of Nd doping (marked as *x* in the general formula), ranging from 0.0 to 0.2 of the molar composition.

## 2. Experiment

Polycrystalline powders of BNFO compounds (with *x* values ranging from 0.00 to 0.20) were synthesized using the sol-gel technique. In the first step, stoichiometric amounts of analytical purity grade Bi(NO$_3$)$_3$·5H$_2$O, Nd(NO$_3$)$_3$·6H$_2$O, and Fe(NO$_3$)$_3$·9H$_2$O were dissolved in polyethylene glycol (10 ml) to obtain solutions with a molar concentration of iron ions of 0.01 mol. Then, 0.02 moles of tartaric acid were added to the solutions and vigorously stirred for 3 hours. Afterwards, solutions were dried at 343 K for 5 hours to obtain a gels. After first drying, they were washed several times with ethyl alcohol and distilled water to remove unreacted starting chemicals. Next, they were dried overnight at 393 K to obtain xerogels, which were subsequently ground into fine powders. Finally, the obtained precursors were annealed in the air atmosphere at 823 K for 4 hours. Samples investigated in the article are denoted by the BNFO acronym, followed by a number signifying the Nd concentration, for example, BNFO10 is a sample with neodymium concentration *x* = 0.1.

X-ray diffraction (XRD) studies were made using an Empyrean powder diffractometer by Malvern Panalytical. For the studies, the Cu $K_\alpha$ radiation was used in the Bragg-Brentano geometry. Low-temperature measurements were carried out employing the PheniX close-cycle helium refrigerator by Oxford Instruments in the 20–300 K temperature range. The sample was mounted on an Al holder with the support of low-temperature ApiezonN grease for thermal contact with the container. The HTK 1200N high-temperature chamber by Anton Paar was used for high-temperature measurements up to 1200 K. The sample was mounted onto the Al$_2$O$_3$ holder. During the measurements, Ar (purity 6N) flow was used to avoid oxygen uptake from the air. For both low- and high-temperature studies the sample stage position was automatically corrected against thermal displacement. The FullProf suite package was used for data evaluation by the Rietveld method [21]. For estimation of the instrumental broadening the calibration sample (NIST 660 – LaB$_6$) was used.

Mössbauer spectroscopy measurements were performed at room temperature using the transmission geometry with the 14.41-keV line of $^{57}$Fe. The RENON MsAa-4 spectrometer was employed with the LND Kr-filled proportional counter and commercial $^{57}$Co(Rh) source kept at room temperature. He-Ne laser-based Michelson-Morley interferometer was used to calibrate a velocity scale. The Mössbauer absorbers were prepared in powder form using 30 mg of Bi$_{1-x}$Nd$_x$FeO$_3$ compounds mixed with a B$_4$C carrier. The absorber thickness (surface density) amounted to 15 mg·cm$^{-2}$ of the investigated material. Spectra were processed by means of the proper applications belonging to the Mosgraf-2009 suite [22] and fitted within standard transmission integral approximation. Isomer shifts are quoted relative to the value of *α*-Fe at room temperature.

The AQUANTA scanning electron microscope (SEM) was utilized to conduct microstructure and morphology observations. The microscope operated at an acceleration voltage ranging from 10 to 20 kV in the high-resolution mode. Furthermore, chemical analysis was performed using an energy-dispersive X-ray spectrometer (EDX) integrated with the microscope. Prior to the SEM/EDX measurement, the samples underwent sputtering with graphite.

Magnetic isothermal loops *M*(*H*) were gathered at room temperature up to 9.5 kOe using the Lake Shore vibrating sample magnetometer Cryotronics Inc. model 7300 (Westerville, OH, USA.)

Differential scanning calorimetry (DSC) was used for specific heat measurements. The samples in powder form with an average weight of 60–80 mg have been placed in the aluminum crucible. The measurements were conducted using a Netzsch DSC F3 Maia calorimeter (Netzsch, Selb, Germany) operating in an argon atmosphere at a flow rate of 40 mL·min$^{-1}$ in the temperature range of 123–773 K with a temperature growth rate of 5 K·min$^{-1}$.

## 3. Results and discussion

### 3.1 Structure

It is well known that the standard solid-state sintering method applied for the preparation of materials based on BiFeO$_3$ usually leads to the formation of significant amounts of secondary phases [23]. This is likely connected with the evaporation of Bi in high-temperature processes. On the other hand, the samples obtained by the sol-gel method and subjected to subsequent isothermal annealing are relatively pure [17,24,25]. Our recent work has revealed that the purest BiFeO$_3$ phase was achieved by annealing the product of the sol-gel route in the temperature range of 723–823 K. [26]. Higher temperatures of the thermal treatment cause enhancement of Bi evaporation and promote the formation of secondary sillenite phases like Bi$_2$Fe$_4$O$_9$, and Bi$_{25}$FeO$_{40}$. That is why we have finally decided to select $T$ = 823 K for annealing our BNFO samples.

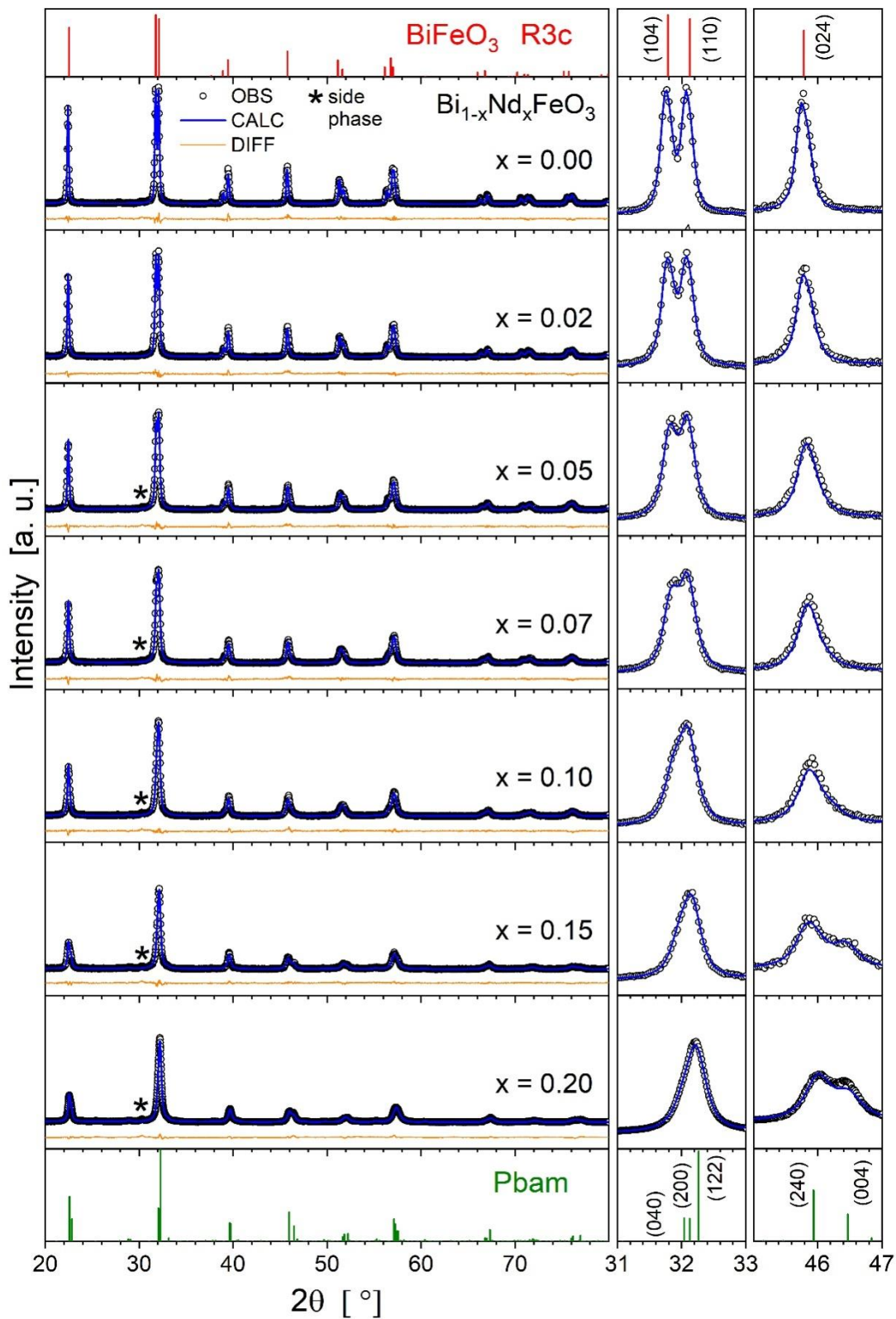

Fig. 1 XRD patterns of BNFO solid solutions. The right panel shows a magnification of peaks in the vicinity of 32° and 46° of 2$\theta$ angle. The standard XRD pattern of BiFeO$_3$ (PDF2 card number 01-082-1254) was included at the top of the figure for comparison. The bottom panel shows the pattern for the *Pbam* phase obtained by Rietveld refinement applied for the BNFO20 sample.

Comprehensive XRD investigations were performed to study the structure of the BNFO samples. Fig. 1 shows the results of the Rietveld refinement procedure applied for XRD data registered for all of the obtained specimens. First of all, high purity of obtained materials was evidenced. In fact, a trace of the most intensive peak for the tiny amount of some secondary phase can be seen close to 30.5° of $2\theta$ angle (marked by a star in Fig.1) The diffractogram recorded for the parent BFO compound agrees well with the standard pattern shown at the top panel of Fig. 1. The best numerical fitting of the diffractogram was achieved assuming rhombohedrally distorted perovskite-like $ABO_3$ structure described by the *R*3*c* space group. Following the evolution of diffractograms for BNFO solid solution along with the increase of *x*, one can note the gradual structural transformation from the *R*3*c* symmetry to orthorhombic *Pbam* structure (see the right panel of Fig. 1). It is worth noting that most of the authors report structural transformation from *R*3*c* to *Pbnm* symmetry [27], while there is only a few works revealing appearance of *Pbam* phase in the case of rare earth-doped $BiFeO_3$ [14,18].

Fig. 2 depicts the decrease of the *R3c* phase fraction with the increasing concentration of Nd ions. Thus, a gradual composition-driven structural transformation from *R*3*c* to *Pbam* symmetry is proved. It is evident that for BNFO10 the contribution of both phases is close to 50% while for BNFO20 the material completely transformed to the *Pbam* structure. Structural data for BFO and BNFO20 samples obtained from the Rietveld refinement procedure are gathered in Tab. 1.

Table 1. Structural data for BFO and BNFO20 samples derived from Rietveld refinement. N denotes the number of formula units per unit cell, x,y,z – fractional atomic coordinates, SOF – site occupation factor.

| Sample | Crystal system Space group | N | Lattice parameters [Å] | Atom positions | x | y | z | SOF |
|---|---|---|---|---|---|---|---|---|
| BFO | rhombohedral *R*3*c* (161) hex. axes | 6 | a=b= 5.5796(4) c = 13.8697(8) | Bi (6a) | 0 | 0 | 0 | 1 |
| | | | | Fe (6a) | 0 | 0 | 0.2215 | 1 |
| | | | | O (18b) | 0.4514 | 0.0174 | 0.9508 | 1 |
| BNFO20 | orthorhombic *Pbam* (55) | 8 | a = 5.5905(4) b = 11.1366(7) c = 7.8153(5) | Bi1 (4g) | 0.2171 | 0.3704 | 0 | 0.8 |
| | | | | Nd1 (4g) | 0.2171 | 0.3704 | 0 | 0.2 |
| | | | | Bi2 (4h) | 0.2344 | 0.3786 | 0.5 | 0.8 |
| | | | | Nd2 (4h) | 0.2344 | 0.3786 | 0.5 | 0.2 |
| | | | | Fe (8i) | 0.2285 | 0.1290 | 0.2373 | 1 |
| | | | | O1 (4g) | 0.7913 | 0.3998 | 0 | 1 |
| | | | | O2 (4h) | 0.7758 | 0.3414 | 0.5 | 1 |
| | | | | O3 (4e) | 0 | 0 | 0.7182 | 1 |
| | | | | O4 (4f) | 0 | 0.5 | 0.6981 | 1 |
| | | | | O5 (8i) | 0.0406 | 0.3057 | 0.2573 | 1 |

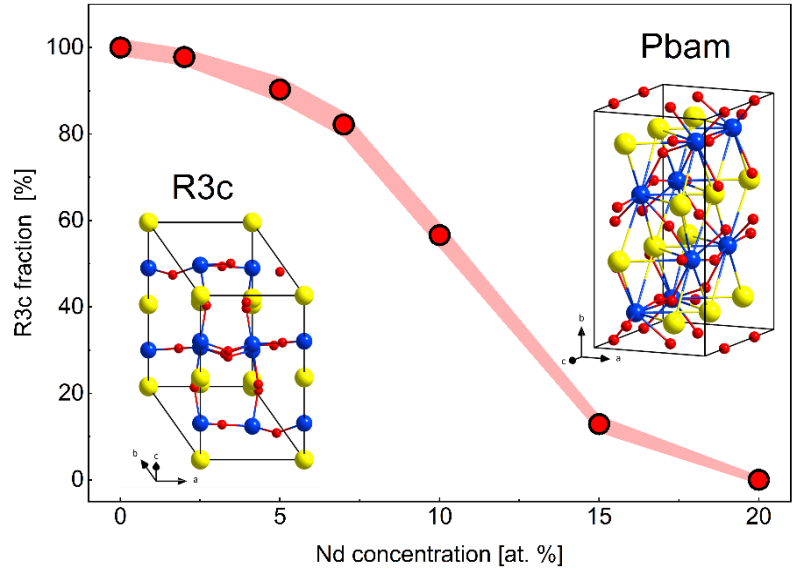

Fig. 2 *R*3*c* phase fraction as a function of increasing concentration of Nd ions. Unit cells of *R*3*c* and *Pbam* were shown as insets. Yellow balls are used for the Bi/Nd sites while the blue and red balls depict the Fe and O sites respectively. The thickness of each "guide to the eye"-line represents uncertainty levels for the dataset.

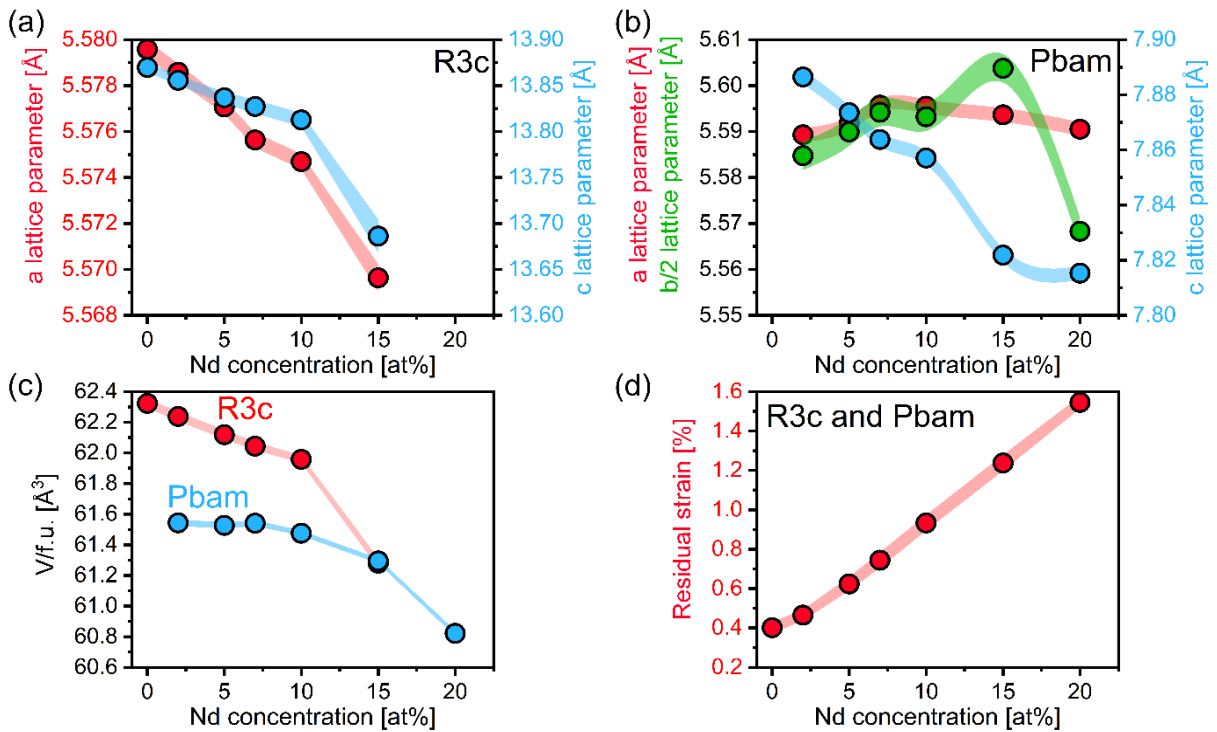

Fig. 3 Lattice parameters derived from Rietveld refinement of XRD data for (a) R3c phase, (b) Pbam phase. The (c) graph depicts unit cell volume per formula unit for both structures, while panel (d) shows residual strain for both phases. The thickness of each "guide to the eye"-line represents uncertainty limits for datasets.

The evolution of structural parameters with the increase of Nd concentration in $Bi_{1-x}Nd_xFeO_3$ solid solution can be seen in Fig. 3. In particular, the drops of *a* and *c* lattice parameters for *R*3*c* phase are visible (Fig. 3a). This is because $Bi^{3+}$ ions (ionic radius $r_{Bi}$ = 1.45 Å for oxygen coordination number $N$ = 12) in BFO crystal lattice were partially substituted by smaller $Nd^{3+}$ ions ($r_{Nd}$ = 1.27 Å) [28] leading to the lattice contraction. The same trend for the *Pbam* phase is less evident. Namely, while there is a significant decrease of the *c* parameter the *a* and *b* parameters seem to initially increase for small concentrations of Nd ions and then decrease for the concentrations close to 20 at. % (Fig. 3b). Following the trends from Fig. 3c one can note a significant reduction of the crystal cell volume per formula unit for the *R*3*c* phase as well as for the *Pbam* structure. Yet, for both phases, the unit cell decreases, which indicates that Nd ions are introduced into both structures. Interestingly, the refinements revealed that residual strain raises significantly with Nd doping, which is understandable, due to the aforementioned differences in the ionic radii of Bi and Nd. It is worth noting that residual strains for samples with x > 10 are relatively high. Apart from strains related to differences in atomic sizes of constituents, the high level of strain can be also associated to a large number of point defects and arising atomic disorder at the 6*a* site for *R*3*c* phase and at the 4*g* or the 4*h* for the *Pbam* phase.

**3.2 Morphology and microstructure**

SEM and EDX observations were performed to study the morphology and microstructure of the BNFO powders. Fig. 4 shows micrographs registered for all of the samples (left panel) and the corresponding distribution of particle size (right panel). As it can be seen the particles are smooth with the shape close to spherical. The distributions are Gaussian-like with the center shifting towards smaller particle size with the increase in Nd concentration. The results of average particle size <*d*> measurements are summarized in Fig. 5a. A Significant drop of <*d*> from 93 nm down to about 35 nm can be observed with the growth of Nd concentration. It is worth noting that the <*d*> less than 62 nm (which is the spin cycloid period) was observed for samples with $x \geq 0.07$. The smallest mean particle size <*d*> = 33.2 nm was revealed for BNFO15 powder.

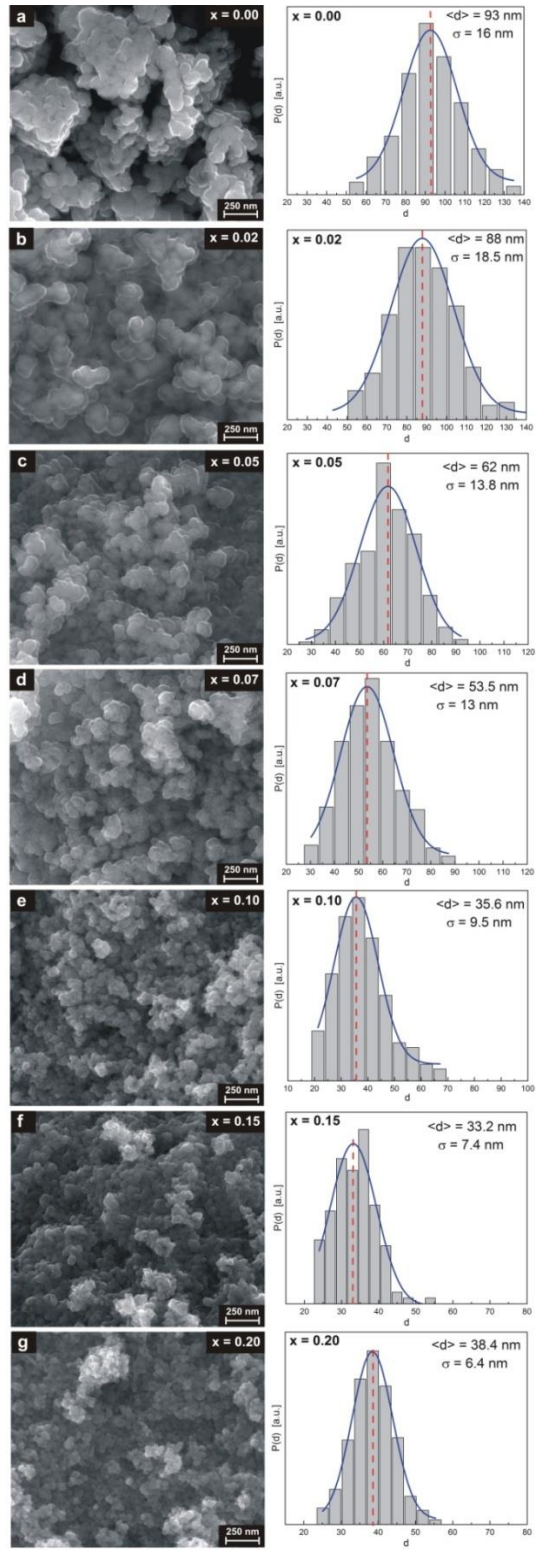

Fig. 4 SEM micrographs recorded for BNFO solid solutions with *x* = 0–0.2 (left panel) and corresponding particle size distributions (right panel). <*d*> denotes the mean particle size whereas *σ* stands for the standard deviation which quantifies the dispersion of each distribution.

EDX studies were carried out for all of the $Bi_{1-x}Nd_xFeO_3$ samples. The derived concentrations of elements were found to be close to the nominal ones (within the limit of experimental uncertainty). Fig. 5b shows the EDX energy spectrum recorded for BNFO10 powder as an example. Additionally, the maps of the concentration of Bi, Nd, and Fe elements were included as insets. One can note no visible segregation of elements which suggests that both the *R3c* and *Pbam* phases exhibit the same chemical composition. Summarizing the XRD and SEM/EDX investigations it can be concluded that disordered solid solutions of BNFO with *R3c* and *Pbam* structure were obtained.

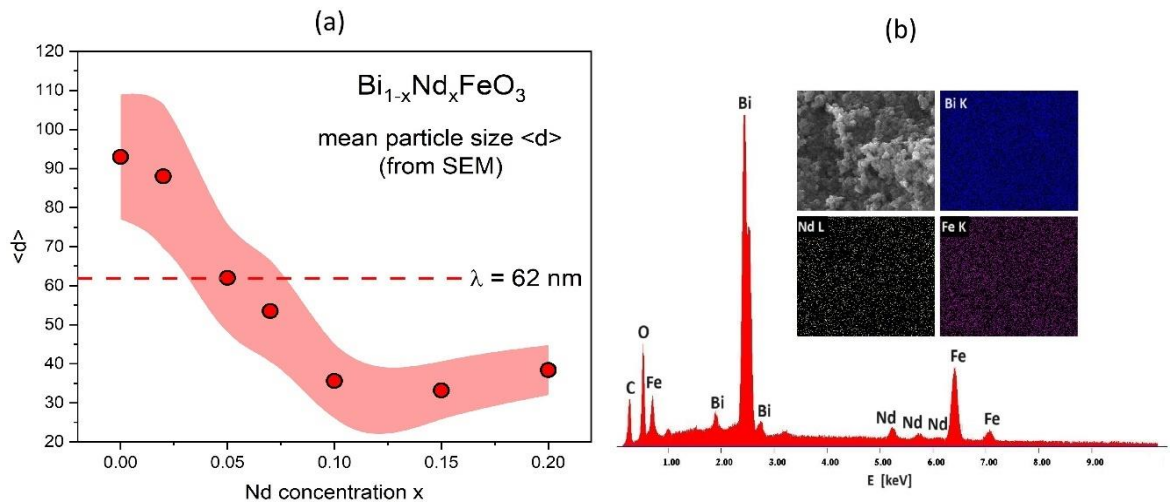

Fig. 5 (a)The dependence of mean particle size <d> measured from high-resolution SEM micrographs as a function of Nd ions concentration, (b) EDX spectrum registered for $Bi_{0.9}Nd_{0.1}FeO_3$ and corresponding maps of Bi, Nd, and Fe elements concentration. The filled area in the left graph represents standard deviation limits for mean particle size.

### 3.3 DSC studies

Fig. 6 presents the results of the specific heat measurements for the BNFO series. AFM-PM phase transition (PT) is visualized as a lambda shape anomaly, occurring near $T_N$ = 640 K for all samples. Wide *λ*-shape anomalies are characteristic of the second-order phase transition. While the addition of Nd ions did not affect the Néel temperature, it should however affect the $T_C$ temperature of the ferro-paraelectric (FE-PE) phase transition. The paper [29] mentioned that the peak corresponding to the ferroelectric transition gradually decreases as x increases, and around the composition *x* = 0.13, $T_C$ and $T_N$ converge. Above x = 0.13, $T_C$ temperature continues to decline. Our measurements did not notice any anomaly related to the ferroelectric phase transition. It may seem unusual but the explanation is simple. The problem is related to the manometer grain size of the ceramics under examination. It is well-known that the ferroelectric effect decreases in ceramics with fine grains below $1\mu m$ [30]. Hence the absence of a visible anomaly associated with the FE-PE phase transition in our measurements is justified.

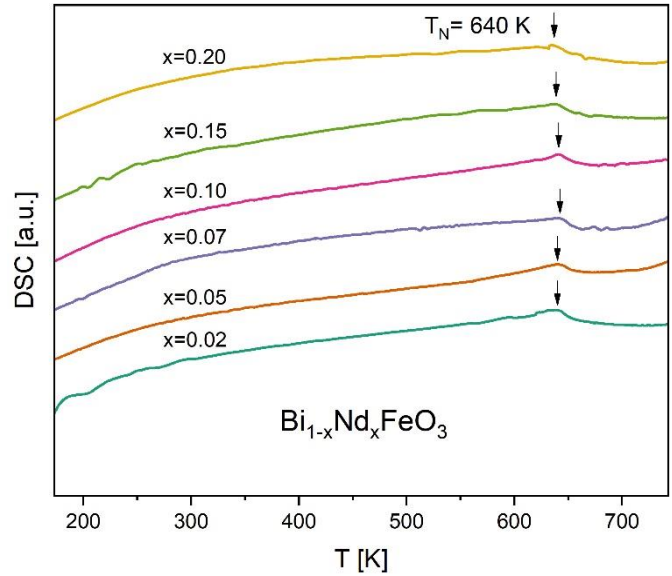

Fig. 6 Temperature dependence of DSC curves for the BNFO series.

### 3.4 Mössbauer spectroscopy studies

To study of the cycloidal spin ordering in $Bi_{1-x}Nd_xFeO_3$, the Mössbauer spectroscopy technique was applied. Fig. 7 shows the obtained spectra and the results of their fitting. It is well established [26,31,32] that cycloidal spin ordering in the sample is manifested in the peculiar shape of Mössbauer spectra. Namely, there is a significant difference in the intensities of lines in twin pairs: 1 and 6, 2 and 5. Moreover, an inhomogeneous broadening of spectral lines is observed. All these features are clearly visible in the case of the pure BFO sample (top spectrum in the left panel of Fig. 7). It can be noted that the asymmetry of spectra gradually diminishes as the content of Nd increases. Finally, for the sample with $x = 0.20$, a symmetric sextet with broadened lines is observed. This time, however, the broadening has a homogeneous character and the spectrum is typical for a single $^{57}Fe$ site in disordered atomic surroundings.

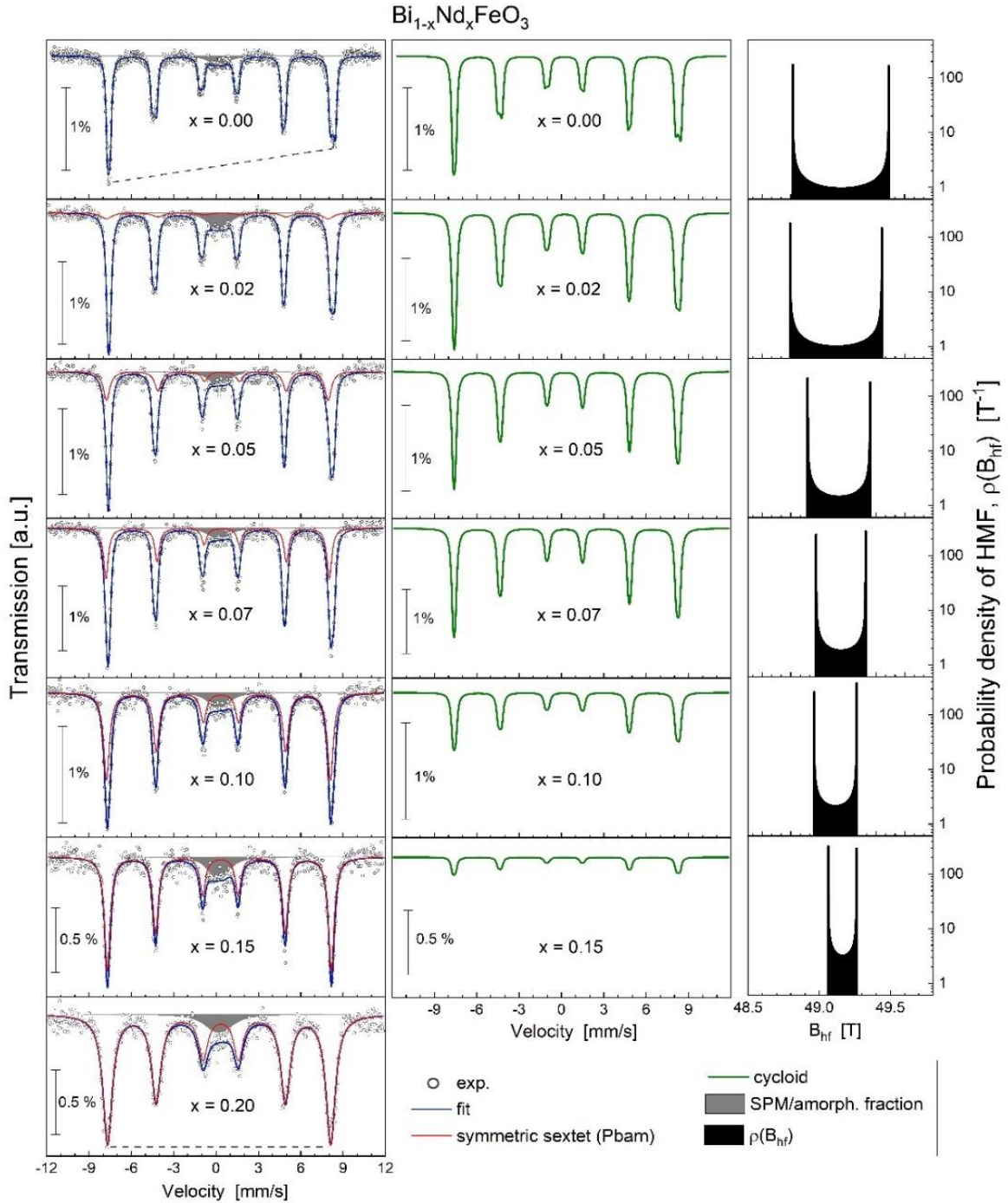

Fig. 7 The Results of Mössbauer spectra fitting for the BNFO series (left panel), the component fitted for cycloid (central panel), and corresponding distributions of hyperfine magnetic field probability density (right panel).

Detailed analyses showed that the Mössbauer spectra consisted of two or three different components, depending on the Nd concentration. The common feature of all spectra is the presence of a broad quadrupolar doublet (marked by a gray color in Fig. 7), which is characteristic of paramagnetic phases. The contribution of the doublet increases with the growth of Nd concentration. This can be attributed to the presence of iron-containing secondary phases (XRD also shows a small signal from the side phase for the highest Nd concentrations) as well as to superparamagnetic relaxation of the smallest fraction of grains or small amounts of amorphous BiFeO$_3$ [26]. It is important to recall that the average particle size derived from the SEM (Fig. 5a) decreased to approximately 30–40 nm for the samples with the highest Nd content. Consequently, the smallest fraction of grains may exhibit superparamagnetic behavior.

The magnetic part of the spectrum for the parent BiFeO$_3$ compound was fitted with the model assumed for spin cycloid described in [33]. It considers two effects: 1) gradual rotation of individual Fe magnetic moments with respect to [001] direction in hexagonal axes and 2) small modulation of the value of hyperfine magnetic field (HMF) along the wave vector of the spin cycloid, i.e. [1-10]$_{hex}$. The latter can be seen as a spin density wave (SDW) with the propagation vector along [1-10]$_{hex}$ direction. According to the model, the angular distribution of HMF can be expressed as:

$$B(\varphi) = B_0 exp\,(P_{20} cos^2 \varphi) \qquad (1)$$

where: $B_0$ is the scaling field related to the low-field boundary of HMF distribution, $\varphi$ - the angle between individual spin (or HMF) and electric polarization direction [001]$_{hex}$, $P_{20}$ quantifies a departure from the circularity of the cycloid i.e. the cycloid ellipticity.

For small values of the $P_{20}$ parameter, the formula can be approximated by:

$$B(\varphi) = B_0 (1 + P_{20} cos^2 \varphi) \qquad (2)$$

Moreover, an axially symmetrical electric field gradient (EFG) at the $^{57}$Fe site was assumed. The main axis of EFG coincides with the electric polarization vector direction, i.e. [001]$_{hex}$ [32]. Thus, the quadrupole shift (QS) parameter was taken constant for all the subspectras producing HMF field distribution.

The magnetic part of the spectra for samples with $x$ = 0.02–0.15 was fitted by a convolution of the cycloidal part (green lines in the central panel of Fig. 7) and symmetric sextet (red color line in the left panel of Fig. 7) ascribed to antiferromagnetic *Pbam* phase. It is worth mentioning that the contributions of both components were kept free during the fitting procedure. The HMF distribution functions corresponding to the fitted cycloidal component are shown in the right panel of Fig. 7. They exhibit the *U*-type shape characteristic of noncolinear magnetic structures or spin density waves [31]. The left arm of the distributions is a peak obtained for the $B_0$ HMF value and corresponds to $\varphi$ = π/2 or $\varphi$ = 3π/2. The right maximum observed for the $B_1$ value is related to $\varphi$ = 0 or $\varphi$ = π cases. Thus, the $B_0$ and $B_1$ fields can be denoted also as $B_\perp$ and $B_{||}$ respectively as it was done in [26,31]. The dependence of $B_0$ and $B_1$ on Nd concentration is shown in Fig. 8b together with the course of HMF value for the symmetric sextet ascribed to the *Pbam* phase. It is apparent that the values of $B_0$ and $B_1$ become similar in values to another as the doping level increases. Fig. 8a shows a comparison of the contribution of the cycloidal part derived from Mössbauer spectra fitting with the contribution of the *R*3*c* phase determined from XRD. They agree well indicating that the disappearance of spin cycloid with the growth of Nd concentration is associated with the progressive structural transformation from *R*3*c* to *Pbam* structure. The values of QS (Fig. 8c) for both the cycloidal component as well as for the symmetric sextet weakly depend on the doping level and they reach 0.5–0.3 mm·s$^{-1}$ and from –0.5 to –0.25 mm·s$^{-1}$ respectively. The decrease in the absolute value of QS can be connected with the increase of the local symmetry of EFG experienced by the $^{57}$Fe nuclear probe. The decrease in the $P_{20}$ parameter (quantifying the cycloid ellipticity for small $P_{20}$ values) with increasing Nd concentration is shown in Fig. 8d suggesting that the increase of $x$ causes a decrease in a spin density wave amplitude (a drop of the width of HMF distribution visible in the right panel of Fig. 7) and the tendency of the cycloid to became more circular (Fig. 8d).

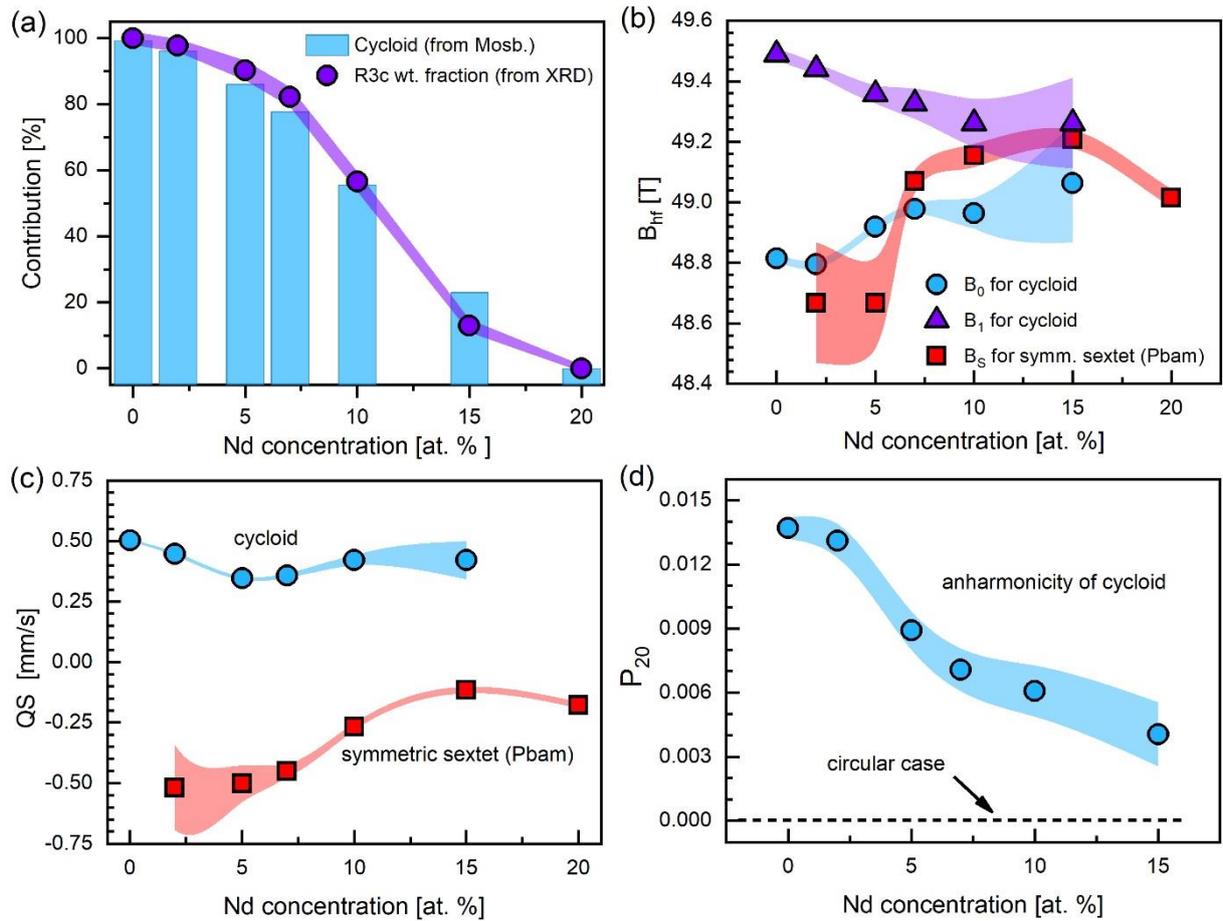

Fig. 8 The parameters derived from Mössbauer spectra fitting as a function of Nd concentration: a) contribution of cycloid from MS compared to the contribution of *Pbam* from XRD, b) hyperfine magnetic field values for cycloid and symmetric sextet, c) quadrupole shift values, d) values of $P_{20}$ anharmonicity parameter. The thickness of each "guide to the eye"-line represents uncertainty limits for datasets.

### 3.4 Magnetization measurements

Magnetization *vs.* magnetic field characteristics for all the studied samples are shown in Fig. 9a. For the parent $BiFeO_3$ compound, a linear, weakly rising trend (red color), typical of a compensated antiferromagnet was observed. This is another confirmation that our $BiFeO_3$ samples, having about 100 nm grain, exhibit bulk-like properties i.e. the spin cycloid averages out a weak magnetization and no hysteresis loop is observed. Doping bismuth ferrite with Nd ions causes the emergence of weak ferromagnetic properties evidenced by an *S*-type hysteresis loop. The magnetization values recorded under $H$ = 9.5 kOe magnetic field for all the studied samples are compared in Fig. 9b. One can note rather slow growth of magnetization in the $x$ = 0–0.1 range. Interestingly, the highest magnetization and the broadest hysteresis loop occur for the sample with $x$ = 0.15. Moving to the sample with $x$ = 0.20 (red color) we can note an abrupt loss of magnetic properties and the value of magnetization very close to zero. This is an unexpected result, as others have reported weak ferromagnetism for the orthorhombic phases of BNFO[14,18]. To explain the trend shown in Fig. 9b two effects occurring upon increasing Nd ions concentration need to be considered; namely a significant drop of grain size (Fig. 5a) and a gradual disappearance of a spin cycloid (Fig. 8a) connected with progressive phase transformation to *Pbam* symmetry. It can be noted that the sample with x = 0.15 exhibits the smallest mean particle size 32(7) nm. This is well below the cycloid period of 62 nm. Thus, the cycloid cannot cancel out the ferromagnetic moment arising from the spin canting. Moreover, a relatively high surface-to-volume ratio can enhance magnetization by uncompensated surface spins of the antiferromagnetic sample.

For the BNFO20 sample, a weak ferromagnetic response was also observed as the *Pbam* structure permits spin canting. In this case, however, the signal is significantly reduced in comparison with BNFO15. This can be explained by a relatively high level of internal strains as evidenced by XRD results (Fig. 3d). Mössbauer studies revealed a symmetric sextet for the BNFO20 proving antiferromagnetic order. However, the spectrum was characterized by the broadest spectral lines (FWHM = 0.70(1) mm/s) proving a high level of structural and atomic disorder. Thus, one can claim the temperature of annealing $T$ = 823 K was not sufficient to complete the diffusion process and to cause total homogenization in the case of the BNFO20 sample.

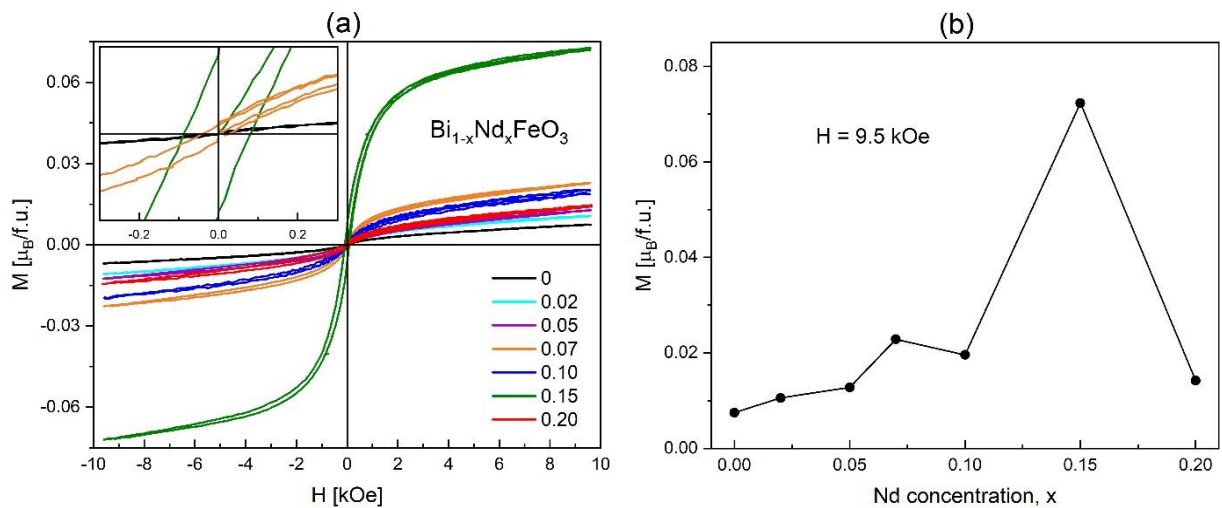

Fig. 9 (a) Magnetization hysteresis loops recorded for BNFO samples and, (b) the dependence of magnetization on the Nd concentration for the applied field $H$ = 9.5 kOe.

**Summary and Conclusions**

In this study, we investigated the effect of Nd dopant concentration on the crystallographic structure, magnetic properties, and cycloidal spin order of the BNFO solid solution. The findings indicate that as the Nd concentration increases, there is a gradual transformation from the *R3c* structure to the *Pbam* orthorhombic phase. This structural change was accompanied by the gradual disappearance of the spin cycloid, what demonstrates that the destruction of the spin cycloid is caused by the structural transformation from the rhombohedral *R3c* to orthorhombic *Pbam* phase. Appearing of the weak ferromagnetic properties, which are often caused by magnetic fields, strains, or doping of BFO, can be associated here to symmetric sextet emerging with the *Pbam* orthorhombic phase lead by increasing the Nd content. The sample with $x$ = 0.15 exhibited the highest magnetization and the broadest magnetic hysteresis loop. This was attributed to the small mean particle size of the particle in the sample, what prevented the complete cancellation of weak ferromagnetism, as well as to uncompensated surface spins. The Néel temperature was found to be insensitive to Nd dopant concentration, remaining at about 640 K for all samples, as revealed DSC studies. Furthermore, the BNFO20 sample exhibited significantly reduced magnetic signal due to high level of strains and structural disorder.

**Acknowledgments**

The work at AGH University was supported by the National Science Centre, Poland, Grant OPUS: 2021/41/B/ST3/03454; the Polish National Agency for Academic Exchange under "Polish Returns 2019" Programme No. PPN/PPO/2019/1/00014; and the "Excellence Initiative-Research University" program for AGH University of Krakow. The work at TU Wien was supported by FWF Project P 35945-N; and the OeAD (Österreichische Austauschdienst)-GmbH Austro-Serbian Project No. RS24/2022.

**Conflict of interest**

On behalf of all authors, the corresponding author states that there is no conflict of interest.